\pdfoutput=1
\documentclass[12pt,a4paper]{article}

\usepackage{ifthen} 
\newboolean{pdflatex}
\setboolean{pdflatex}{true} 

\newboolean{articletitles}
\setboolean{articletitles}{true} 

\newboolean{uprightparticles}
\setboolean{uprightparticles}{false} 

\def\paperauthors{LHCb collaboration} 
\def\paperasciititle{Precision measurement of the Xib0 baryon lifetime} 
\def\papertitle{Precision measurement of the \newline $\Xibz$ baryon lifetime} 
\def\paperkeywords{{High Energy Physics}, {LHCb}} 
\def\papercopyright{\the\year\ CERN for the benefit of the LHCb collaboration} 
\def\paperlicence{CC BY 4.0 licence}
\def\paperlicenceurl{https://creativecommons.org/licenses/by/4.0/}

\newif\ifEnableSectionTOCLinks
\EnableSectionTOCLinksfalse 

\usepackage[top=1in, bottom=1.25in, left=1in, right=1in]{geometry}

\usepackage{microtype}
\usepackage{lineno}  
\usepackage{xspace} 
\usepackage{caption} 

\usepackage{graphicx}  
\usepackage{color}
\usepackage{colortbl}
\graphicspath{{./figs/}} 

\usepackage{amsmath} 
\usepackage{amssymb}
\usepackage{amsfonts}
\usepackage{upgreek} 

\newcommand*\patchAmsMathEnvironmentForLineno[1]{%
\expandafter\let\csname old#1\expandafter\endcsname\csname #1\endcsname
\expandafter\let\csname oldend#1\expandafter\endcsname\csname
end#1\endcsname
 \renewenvironment{#1}%
   {\linenomath\csname old#1\endcsname}%
   {\csname oldend#1\endcsname\endlinenomath}%
}
\newcommand*\patchBothAmsMathEnvironmentsForLineno[1]{%
  \patchAmsMathEnvironmentForLineno{#1}%
  \patchAmsMathEnvironmentForLineno{#1*}%
}
\AtBeginDocument{%
\patchBothAmsMathEnvironmentsForLineno{equation}%
\patchBothAmsMathEnvironmentsForLineno{align}%
\patchBothAmsMathEnvironmentsForLineno{flalign}%
\patchBothAmsMathEnvironmentsForLineno{alignat}%
\patchBothAmsMathEnvironmentsForLineno{gather}%
\patchBothAmsMathEnvironmentsForLineno{multline}%
\patchBothAmsMathEnvironmentsForLineno{eqnarray}%
}

\usepackage[pdftex,
            pdfauthor={\paperauthors},
            pdftitle={\paperasciititle},
            pdfkeywords={\paperkeywords}]{hyperref}
\usepackage{hyperxmp}
\hypersetup{
    pdfcopyright={Copyright (C) \papercopyright},
    pdflicenseurl={\paperlicenceurl}
}

\usepackage[colorinlistoftodos,textsize=scriptsize]{todonotes}

\usepackage[bottom,flushmargin,hang,multiple]{footmisc}

\usepackage[all]{hypcap} 

\usepackage{xspace} 
\usepackage{upgreek}

\def\lhcb   {\mbox{LHCb}\xspace}

\def\MagUp {\mbox{\em Mag\kern -0.05em Up}\xspace}

\ifthenelse{\boolean{uprightparticles}}%
{

 \def\Ppi         {\ensuremath{\uppi}\xspace}

 \def\PDelta      {\ensuremath{\Delta}\xspace}                 
 \def\PXi         {\ensuremath{\Xi}\xspace}                 
 \def\PLambda     {\ensuremath{\Lambda}\xspace}                 
 \def\PSigma      {\ensuremath{\Sigma}\xspace}                 
 \def\POmega      {\ensuremath{\Omega}\xspace}                 
 \def\PUpsilon    {\ensuremath{\Upsilon}\xspace}
 \let\oldPi\Pi
 \def\PPi         {\ensuremath{\oldPi}\xspace}

 \def\PB      {\ensuremath{\mathrm{B}}\xspace}                 
                  
 \def\PD      {\ensuremath{\mathrm{D}}\xspace}

 \def\PK      {\ensuremath{\mathrm{K}}\xspace}

 \def\Pb      {\ensuremath{\mathrm{b}}\xspace}                 
 \def\Pc      {\ensuremath{\mathrm{c}}\xspace}                 
 \def\Pd      {\ensuremath{\mathrm{d}}\xspace}

 \def\Pi      {\ensuremath{\mathrm{i}}\xspace}

 \def\Ps      {\ensuremath{\mathrm{s}}\xspace}                 
                  
 \def\Pu      {\ensuremath{\mathrm{u}}\xspace}

 \def\thebaroffset{0.0em}
}
{

 \def\Ppi         {\ensuremath{\pi}\xspace}

 \mathchardef\PDelta="7101
 \mathchardef\PXi="7104
 \mathchardef\PLambda="7103
 \mathchardef\PSigma="7106
 \mathchardef\POmega="710A
 \mathchardef\PUpsilon="7107
 \mathchardef\PPi="7105
                  
 \def\PB      {\ensuremath{B}\xspace}                 
                  
 \def\PD      {\ensuremath{D}\xspace}

 \def\PK      {\ensuremath{K}\xspace}

 \def\Pb      {\ensuremath{b}\xspace}                 
 \def\Pc      {\ensuremath{c}\xspace}                 
 \def\Pd      {\ensuremath{d}\xspace}

 \def\Pi      {\ensuremath{i}\xspace}

 \def\Ps      {\ensuremath{s}\xspace}                 
                  
 \def\Pu      {\ensuremath{u}\xspace}

 \def\thebaroffset{0.18em}
}
\newcommand{\offsetoverline}[2][\thebaroffset]{\kern #1\overline{\kern -#1 #2}}%

\makeatletter
\ifcase \@ptsize \relax
  \newcommand{\miniscule}{\@setfontsize\miniscule{4}{5}}
\or
  \newcommand{\miniscule}{\@setfontsize\miniscule{5}{6}}
\or
  \newcommand{\miniscule}{\@setfontsize\miniscule{5}{6}}
\fi
\makeatother

\DeclareRobustCommand{\optbar}[1]{\shortstack{{\miniscule (\rule[.5ex]{1.25em}{.18mm})}
  \\ [-.7ex] $#1$}}

\def\uquark    {{\ensuremath{\Pu}}\xspace}
\def\uquarkbar {{\ensuremath{\overline \uquark}}\xspace}

\def\dquark    {{\ensuremath{\Pd}}\xspace}
\def\dquarkbar {{\ensuremath{\overline \dquark}}\xspace}

\def\squark    {{\ensuremath{\Ps}}\xspace}
\def\squarkbar {{\ensuremath{\overline \squark}}\xspace}

\def\cquark    {{\ensuremath{\Pc}}\xspace}

\def\bquark    {{\ensuremath{\Pb}}\xspace}
\def\bquarkbar {{\ensuremath{\overline \bquark}}\xspace}
\def\bbbar     {{\ensuremath{\bquark\bquarkbar}}\xspace}

\def\pion   {{\ensuremath{\Ppi}}\xspace}

\def\pip    {{\ensuremath{\pion^+}}\xspace}
\def\pim    {{\ensuremath{\pion^-}}\xspace}

\def\kaon    {{\ensuremath{\PK}}\xspace}

\def\KorKbar {\kern \thebaroffset\optbar{\kern -\thebaroffset \PK}{}\xspace}

\def\Kp      {{\ensuremath{\kaon^+}}\xspace}
\def\Km      {{\ensuremath{\kaon^-}}\xspace}

\def\D       {{\ensuremath{\PD}}\xspace}

\def\DorDbar {\kern \thebaroffset\optbar{\kern -\thebaroffset \PD}\xspace}
\def\Dz      {{\ensuremath{\D^0}}\xspace}

\def\Dp      {{\ensuremath{\D^+}}\xspace}
\def\Dm      {{\ensuremath{\D^-}}\xspace}

\def\DpDm    {\ensuremath{\Dp {\kern -0.16em \Dm}}\xspace}

\def\Dstarp  {{\ensuremath{\D^{*+}}}\xspace}

\def\B       {{\ensuremath{\PB}}\xspace}

\def\BorBbar {\kern \thebaroffset\optbar{\kern -\thebaroffset \PB}\xspace}

\def\Bd      {{\ensuremath{\B^0}}\xspace}

\def\BdorBdbar {\kern \thebaroffset\optbar{\kern -\thebaroffset \Bd}\xspace}

\def\Bs      {{\ensuremath{\B^0_\squark}}\xspace}

\def\BsorBsbar {\kern \thebaroffset\optbar{\kern -\thebaroffset \Bs}\xspace}

\def\Y#1S{\ensuremath{\PUpsilon{(#1S)}}\xspace}

\def\Lz          {{\ensuremath{\PLambda}}\xspace}

\def\LorLbar     {\kern \thebaroffset\optbar{\kern -\thebaroffset \PLambda}\xspace}

\def\Xires       {{\ensuremath{\PXi}}\xspace}

\def\Omegares    {{\ensuremath{\POmega}}\xspace}

\def\Lc          {{\ensuremath{\Lz^+_\cquark}}\xspace}

\def\Xic         {{\ensuremath{\Xires_\cquark}}\xspace}

\def\Xicp        {{\ensuremath{\Xires^+_\cquark}}\xspace}
\def\Xicprime    {{\ensuremath{\Xires^{'+}_\cquark}}\xspace}

\def\Lb           {{\ensuremath{\Lz^0_\bquark}}\xspace}

\def\Xib          {{\ensuremath{\Xires_\bquark}}\xspace}
\def\Xibz         {{\ensuremath{\Xires^0_\bquark}}\xspace}
\def\Xibm         {{\ensuremath{\Xires^-_\bquark}}\xspace}

\def\Omegab       {{\ensuremath{\Omegares^-_\bquark}}\xspace}

\def\BF         {{\ensuremath{\mathcal{B}}}\xspace}
\def\BR         {\BF}

\def\to                 {\ensuremath{\rightarrow}\xspace}

\def\eff   {{\ensuremath{\varepsilon}}\xspace}

\def\AT#1     {\ensuremath{A_{\mathrm{T}}^{#1}}\xspace}           

\def\C#1      {\ensuremath{\mathcal{C}_{#1}}\xspace}                       
\def\Cp#1     {\ensuremath{\mathcal{C}_{#1}^{'}}\xspace}                    
\def\Ceff#1   {\ensuremath{\mathcal{C}_{#1}^{\mathrm{(eff)}}}\xspace}        
\def\Cpeff#1  {\ensuremath{\mathcal{C}_{#1}^{'\mathrm{(eff)}}}\xspace}       
\def\Ope#1    {\ensuremath{\mathcal{O}_{#1}}\xspace}                       
\def\Opep#1   {\ensuremath{\mathcal{O}_{#1}^{'}}\xspace}                    

\newcommand{\nospaceunit}[1]{\ensuremath{\text{#1}}}       
\newcommand{\aunit}[1]{\ensuremath{\text{\,#1}}}       

\newcommand{\tev}{\aunit{Te\kern -0.1em V}\xspace}
\newcommand{\gev}{\aunit{Ge\kern -0.1em V}\xspace}
\newcommand{\mev}{\aunit{Me\kern -0.1em V}\xspace}
\newcommand{\kev}{\aunit{ke\kern -0.1em V}\xspace}
\newcommand{\ev}{\aunit{e\kern -0.1em V}\xspace}
 
\newcommand{\mevc}{\ensuremath{\aunit{Me\kern -0.1em V\!/}c}\xspace}
\newcommand{\gevc}{\ensuremath{\aunit{Ge\kern -0.1em V\!/}c}\xspace}
\newcommand{\mevcc}{\ensuremath{\aunit{Me\kern -0.1em V\!/}c^2}\xspace}
\newcommand{\gevcc}{\ensuremath{\aunit{Ge\kern -0.1em V\!/}c^2}\xspace}

\def\mum  {\ensuremath{\,\upmu\nospaceunit{m}}\xspace}

\def\fb   {\ensuremath{\aunit{fb}}\xspace}
\def\invfb   {\ensuremath{\fb^{-1}}\xspace}

\def\ps   {\ensuremath{\aunit{ps}}\xspace}

\newcommand{\chisq}{\ensuremath{\chi^2}\xspace}

\newcommand{\chisqip}{\ensuremath{\chi^2_{\text{IP}}}\xspace}

\def\gsim{{~\raise.15em\hbox{$>$}\kern-.85em
          \lower.35em\hbox{$\sim$}~}\xspace}
\def\lsim{{~\raise.15em\hbox{$<$}\kern-.85em
          \lower.35em\hbox{$\sim$}~}\xspace}

\def\pt         {\ensuremath{p_{\mathrm{T}}}\xspace}

\def\ptot       {\ensuremath{p}\xspace}

\def\evtgen     {\mbox{\textsc{EvtGen}}\xspace}

\def\geant      {\mbox{\textsc{Geant4}}\xspace}

\def\photos     {\mbox{\textsc{Photos}}\xspace}

\def\pythia     {\mbox{\textsc{Pythia}}\xspace}

\def\tell1  {TELL1\xspace}
\def\ukl1   {UKL1\xspace}

\newcommand{\lhcborcid}[1]{\href{https://orcid.org/#1}{\hspace*{0.1em}\raisebox{-0.45ex}{\includegraphics[width=1em]{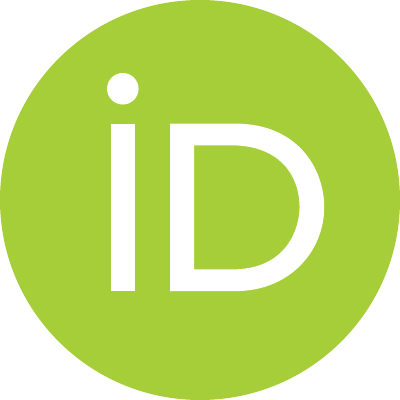}}}}

\hypersetup{
  colorlinks   = true, 
  urlcolor     = blue, 
  linkcolor    = blue, 
  citecolor    = red   
}

\ifEnableSectionTOCLinks
    \usepackage[explicit]{titlesec} 
    
    \let\oldcontentsline\contentsline
    \renewcommand

    \titleformat{\section}{\normalfont\Large\bf}{\hyperlink{tocsection.\thesection}{{\thesection} \parbox[t]{\dimexpr\textwidth-1pc}{#1}}}{1pc}{}

    \titleformat{\subsection}{\normalfont\bf}{\hyperlink{tocsubsection.\thesubsection}{{\thesubsection} \parbox[t]{\dimexpr\textwidth-1pc}{#1}}}{1pc}{}

    \titleformat{name=\section,numberless}[display]{}{}{0pt}{\normalfont\Huge\bfseries #1}
\fi

\usepackage{cite} 
\usepackage{mciteplus}

\usepackage{longtable} 

\begin{document}

\renewcommand{\thefootnote}{\fnsymbol{footnote}}
\setcounter{footnote}{1}


\begin{titlepage}
\pagenumbering{roman}

\vspace*{-1.5cm}
\centerline{\large EUROPEAN ORGANIZATION FOR NUCLEAR RESEARCH (CERN)}
\vspace*{1.5cm}
\noindent
\begin{tabular*}{\linewidth}{lc@{\extracolsep{\fill}}r@{\extracolsep{0pt}}}
\ifthenelse{\boolean{pdflatex}}
{\vspace*{-1.5cm}\mbox{\!\!\!\includegraphics[width=.14\textwidth]{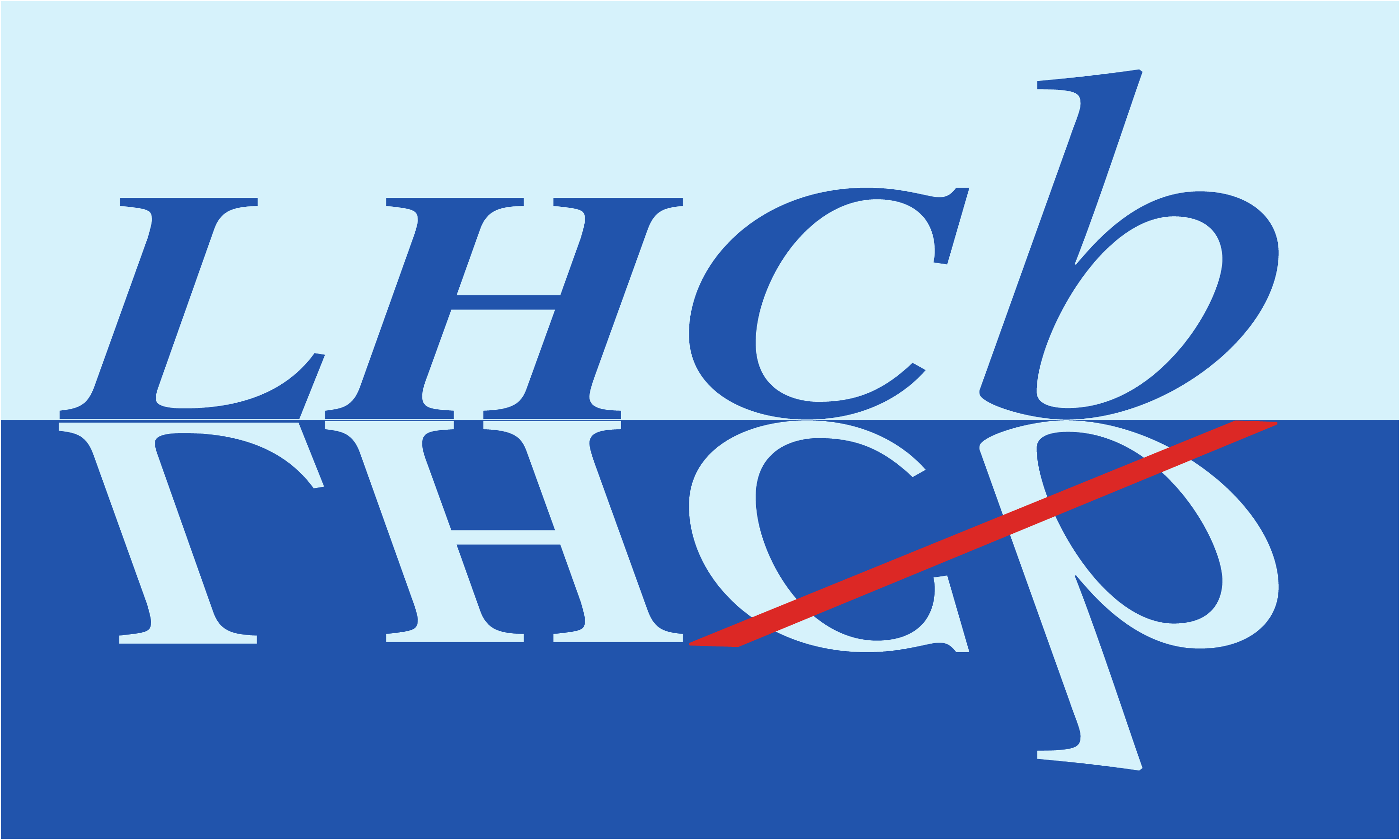}} & &}%
{\vspace*{-1.2cm}\mbox{\!\!\!\includegraphics[width=.12\textwidth]{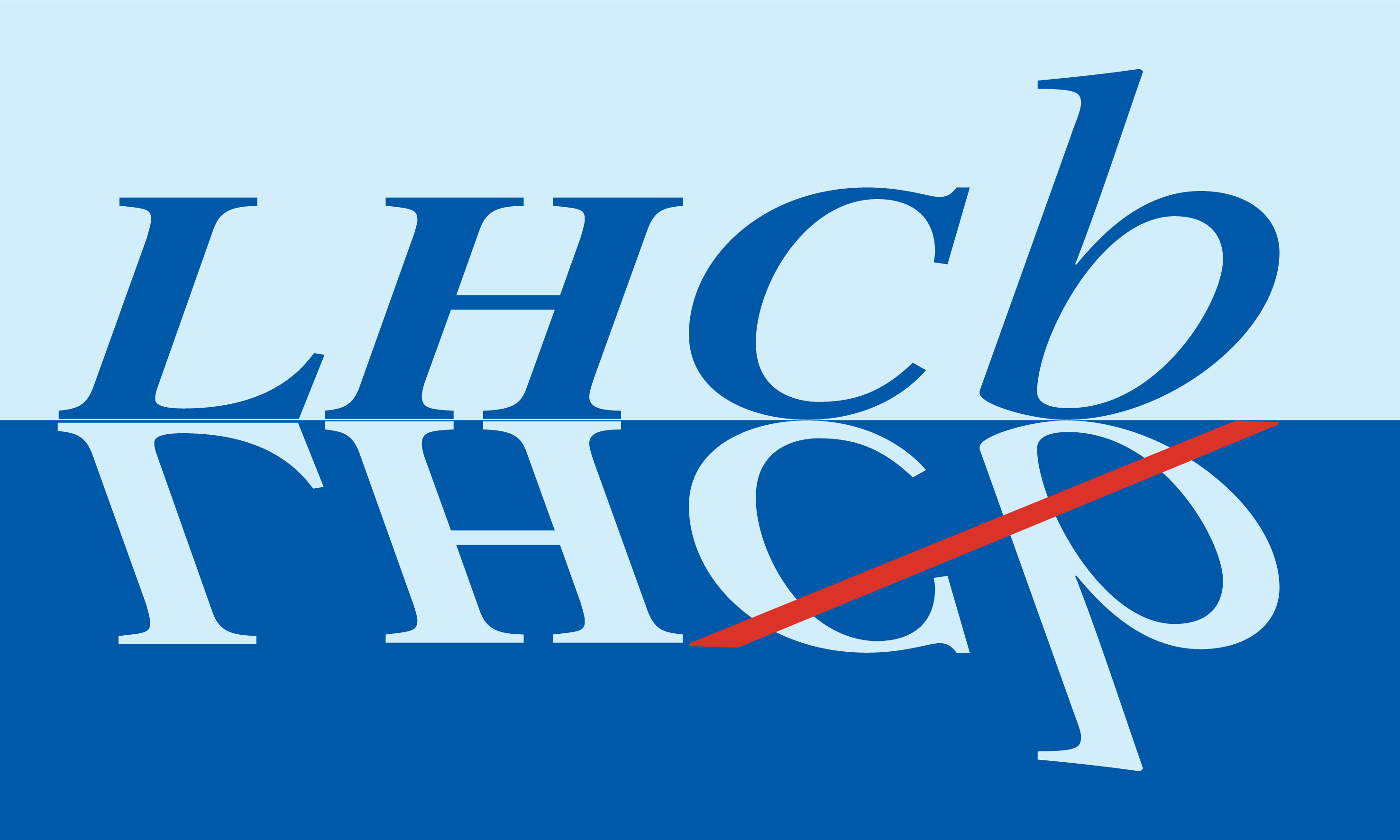}} & &}%
\\
 & & CERN-EP-2025-139 \\  
 & & LHCb-PAPER-2025-023 \\  
 & & Sept. 30, 2025 \\ 
 & & \\
\end{tabular*}

\vspace*{4.0cm}

{\normalfont\bfseries\boldmath\huge
\begin{center}
  \papertitle 
\end{center}
}

\vspace*{2.0cm}

\begin{center}
\paperauthors\footnote{Authors are listed at the end of this paper.}
\end{center}

\vspace{\fill}

\begin{abstract}
  \noindent
A sample of $pp$ collision data, corresponding to an integrated luminosity of 5.4$\invfb$ and collected by the LHCb experiment during LHC Run 2, is used to measure the ratio of the lifetime of the ${\it{\Xi}}_b^0$ baryon to that of the ${\it{\Lambda}}_b^0$ baryon, $r_{\tau}\equiv\tau_{{\it{\Xi}}_b^0}/\tau_{{\it{\Lambda}}_b^0}$. The value ${r_{\tau}^{\rm Run\,2}=1.004\pm0.009\pm0.006}$ is obtained, where the first uncertainty is statistical and the second systematic. This value is averaged with the corresponding value from Run 1 to obtain 
${r_{\tau} = 1.004\pm0.008\pm0.005}$. Multiplying by the known value of the ${\it{\Lambda}}_b^0$ lifetime yields ${{\tau_{{\it{\Xi}}_b^0}} = 1.475\pm0.012\pm0.008\pm0.009~{\rm ps}}$, where the last uncertainty is due to the limited knowledge of the ${\it{\Lambda}}_b^0$ lifetime. This measurement improves the precision of the current world average of the ${\it{\Xi}}_b^0$  lifetime by about a factor of two, and is in good agreement with the most recent theoretical predictions.
\end{abstract}

\vspace*{2.0cm}

\begin{center}
Published in Phys. Rev. D112 (2025) 052012
\end{center}

\vspace{\fill}

{\footnotesize 
\centerline{\copyright~\papercopyright. \href{\paperlicenceurl}{\paperlicence}.}}
\vspace*{2mm}

\end{titlepage}


\newpage
\setcounter{page}{2}
\mbox{~}


\renewcommand{\thefootnote}{\arabic{footnote}}
\setcounter{footnote}{0}


\cleardoublepage


\pagestyle{plain} 
\setcounter{page}{1}
\pagenumbering{arabic}


\section{Introduction}
\label{sec:Introduction}

The heavy-quark expansion
(HQE)~\cite{Lenz:2014jha} is a theoretical framework that predicts the inclusive decay rates of beauty hadrons through an expansion in powers of the strong coupling constant, $\alpha_s$, and $\Lambda_{\rm QCD}/m_b$. Here, $\Lambda_{\rm QCD}$ is the energy scale at which the strong-interaction coupling becomes large, ${\mathcal{O}}(100\mev)$, and $m_b$ is the $\bquark$-quark mass. When combined with precise measurements of heavy-quark decays, the HQE framework can be used to calculate $\bquark$-hadron parameters required for the determination of Cabibbo--Kobayashi--Maskawa~\cite{Cabibbo:1963yz,*Kobayashi:1973fv} matrix elements, which in turn provide constraints on physics beyond the Standard Model.

A stringent test of the HQE framework is to confront its predictions for the total decay widths $\Gamma$ of various $\bquark$ hadrons with precise  measurements of the corresponding lifetimes, $\tau=\hbar/\Gamma$. Here, $\bquark$ hadron refers to a baryon (meson) containing a single $\bquark$ quark and any pair of $u,~d$ or $s$ quarks (a single $\uquarkbar$, $\dquarkbar$ or $\squarkbar$ antiquark). At leading order within the HQE framework, the decay width of all $\bquark$ hadrons is equal to that of the free $\bquark$ quark. Higher-order corrections account for nonperturbative effects, which are described by the kinetic, Darwin and chromomagnetic operators~\cite{Gratrex,Dassinger_2007,SumRules,HigherOrderCorr}, as well as interactions of the $\bquark$ quark with the light valence quarks. It is in the interactions of the $\bquark$ quark with the light valence quarks that significant differences in $\bquark$-hadron lifetimes emerge. The beauty baryons, $\Lb~(bud)$, $\Xibz~(bus)$, $\Xibm~(bds)$ and $\Omegab~(bss)$ have different pairs of light quarks, and thus the measurement of their lifetimes  probes the effects of the light valence quark contributions on their total decay widths. 

A recent precise measurement of the lifetime ratio \mbox{$\tau_{\Xibm}/\tau_{\Lb}=1.078\pm0.012\pm0.007$}~\cite{LHCb-PAPER-2024-010} shows that the light valence quark effects result in a difference of about $8\%$  in the total decay widths of the $\Lb$ and $\Xibm$ baryons, which is in good agreement with the HQE prediction~\cite{Gratrex}. The only measurement of the $\Xibz$ baryon lifetime has been performed by the LHCb collaboration using data samples corresponding to an integrated luminosity of 3\invfb collected at $\sqrt{s}=7$ and 8\tev (Run 1). The
value of the lifetime ratio \mbox{$\tau_{\Xibz}/\tau_{\Lb}=1.006\pm0.018\pm0.010$}~\cite{LHCb-PAPER-2014-021} is reported, which is consistent with the HQE prediction.

In this paper, a new measurement of the $\Xibz$ baryon lifetime is reported using a $pp$ collision data sample collected between 2016 and 2018 (Run\,2) by the LHCb experiment at $\sqrt{s}=13\tev$, corresponding to an integrated luminosity of 5.4\invfb. The integrated luminosity and $\bbbar$ production cross-section are both about a factor of two larger compared to the previous analysis~\cite{LHCb-PAPER-2014-021}. The lifetime ratio $\tau_{\Xibz}/\tau_{\Lb}$ is measured using the decays \mbox{$\Xibz\to\Xicp\pim$} and \mbox{$\Lb\to\Lc\pim$}, where both the $\Xicp$ and $\Lc$ baryons are reconstructed in the $p\Km\pip$ final state. Although the $\Xicp\to p\Km\pip$ decay is Cabibbo-suppressed, the chosen $\Xibz$ decay mode provides the largest signal yield per unit luminosity in LHCb due to the relatively low multiplicity and the absence of long-lived hyperons in the final state.\footnote{The inclusion of charge-conjugate processes is implied throughout.}

The quantity that is measured experimentally is the ratio of efficiency-corrected signal yields as a function of the decay time, $t$,
\begin{align}
    R(t) \equiv\frac{N[\Xibz\to\Xicp\pim](t)}{N[\Lb\to\Lc\pim](t)} \cdot \frac{\eff[\Lb\to\Lc\pim](t)}{\eff[\Xibz\to\Xicp\pim](t)}=R_0\exp{(\lambda t)},
\label{eq:RT}
\end{align}
\noindent where $R_0$ is an overall normalization factor, and $N$ and $\eff$ represent the signal yields and efficiencies, respectively, of the decay mode indicated inside the brackets. The parameter~$\lambda$ is related to the lifetimes of the $\Lb$ and $\Xibz$ baryons through
\begin{align}
    \lambda\equiv\frac{1}{\tau_{\Lb}}-\frac{1}{\tau_{\Xibz}}.
\end{align}
\noindent The ratio of lifetimes is then
\begin{align}
    r_{\tau}\equiv\frac{\tau_{\Xibz}}{\tau_{\Lb}}=\frac{1}{1-\lambda\,\tau_{\Lb}}.
\label{eq:rtau}
\end{align}

\section{Detector and simulation}
The \lhcb detector~\cite{LHCb-DP-2008-001,LHCb-DP-2014-002} is a single-arm forward
spectrometer covering the \mbox{pseudorapidity} range $2<\eta <5$, designed for the study of particles containing \bquark or \cquark
quarks. The detector used for this analysis includes a high-precision tracking system
consisting of a silicon-strip vertex detector surrounding the $pp$
interaction region~\cite{LHCb-DP-2014-001}, a large-area silicon-strip detector located
upstream of a dipole magnet with a bending power of about
$4{\mathrm{\,T\,m}}$, and three stations of silicon-strip detectors and straw
drift tubes~\cite{LHCb-DP-2017-001} placed downstream of the magnet.
The tracking system provides a measurement of the momentum, \ptot, of charged particles with
a relative uncertainty that varies from 0.5\% at low momentum to 1.0\% at 200\gevc. The polarity of the LHCb magnet is alternated regularly throughout each period of data taking. The minimum distance of a track to a primary $pp$ collision vertex (PV), the impact parameter (IP), 
is measured with a resolution of approximately $\sigma_{\rm IP}=(15+29/\pt)\mum$,
where \pt is the component of the momentum transverse to the beam, in\,\gevc.
Different types of charged hadrons are distinguished using information
from two ring-imaging Cherenkov detectors~\cite{LHCb-DP-2012-003}. 
Photons, electrons and hadrons are identified by a calorimeter system consisting of
scintillating-pad and preshower detectors, an electromagnetic
and a hadronic calorimeter. Muons are identified by a
system composed of alternating layers of iron and multiwire
proportional chambers~\cite{LHCb-DP-2012-002}.
The online event selection is performed by a trigger~\cite{LHCb-DP-2019-001},  which consists of a hardware stage (L0), based on information from the calorimeter and muon systems, followed by a software stage, which applies a full event reconstruction. The software stage employs a multivariate algorithm~\cite{BBDT,LHCb-PROC-2015-018} to identify secondary vertices consistent with the decay of a \bquark hadron.

Simulation is required to model the effects of the detector acceptance and the imposed selection requirements. In the simulation, $pp$ collisions are generated using \pythia~\cite{Sjostrand:2007gs,*Sjostrand:2006za} 
with a specific \lhcb configuration~\cite{LHCb-PROC-2010-056}. Decays of unstable particles are described by \evtgen~\cite{Lange:2001uf}, in which final-state radiation is generated using \photos~\cite{davidson2015photos}. The interaction of the generated particles with the detector, and its response, are implemented using the \geant toolkit~\cite{Allison:2006ve, *Agostinelli:2002hh} as described in Ref.~\cite{LHCb-PROC-2011-006}. The underlying $pp$ interaction is reused multiple times, with an independently generated signal decay each time~\cite{LHCb-DP-2018-004}. 
Simulated $\Xibz\to\Xicp\pim$ and $\Lb\to\Lc\pim$ decays are generated separately for each of the three data-taking years: 2016, 2017 and 2018. The simulation reflects the running conditions specific to each year. These conditions mainly account for adjustments in the hardware and software trigger thresholds, which are modified to accommodate variations in the instantaneous luminosity.

\section{Candidate selection}
\label{sec:selection}

Signal $\Xibz$ ($\Lb$) candidates are formed by fitting a $\Xicp\to p\Km\pip$ ($\Lc\to p\Km\pip$) candidate decay with a $\pim$ candidate to a common vertex. Hereafter, the notation $H_b$ and $H_c$ refers to the $\bquark$ and $\cquark$ baryons. All final-state particles from the $H_b$ baryons are required to have trajectories that are significantly detached from all PVs in the event, characterized by the quantity $\chisqip$, which 
is defined as the difference in the vertex-fit $\chisq$ of a given PV reconstructed with and without the particle in question. The final-state hadrons are also required to have particle identification (PID) responses that are consistent with the decay hypothesis.
The $H_c$ candidates are required to be well separated from all PVs in the event, have good decay-vertex fit quality, and have  mass, $M(p\Km\pip)$, within 15\mevcc~(25\mevcc) of the known $\Xicp$ ($\Lc$) baryon mass $m_{\Xicp}$ ($m_{\Lc}$)~\cite{PDG2024}, which corresponds to about 2.5 (3.5) times the mass resolution. The $H_c$ decay time must satisfy \mbox{$-0.2 < t_{\Xicp}< 5.0\ps$} \mbox{($-0.5 < t_{\Lc}< 2.5\ps$)}, where the lower bound accounts for the decay-time resolution, and the upper bound is approximately ten times the known lifetime~\cite{PDG2024}. The decay time of the $H_c$ ($H_b$) baryon is obtained using $t=d\frac{M}{p}$, where $d$, $M$ and $p$ are the measured decay length, mass and total momentum of the $H_c$ ($H_b$) baryon. The $H_b$ decay vertex is required to have a good fit quality and be well separated from all PVs in the event. The PV with the smallest $\chisqip$ is assigned as the production point of the $H_b$ candidate from which the decay time is measured. 

Mass vetoes, in combination with additional PID requirements, are used to suppress cross-feeds from misidentified $D^{+}_{(s)}\to\Kp\Km\pip$, \mbox{$D^{*+}\to\Dz(\Kp\Km)\pip$}, and $\Dp\to\Km\pip\pip$ decays mimicking $\Lc\to p\Km\pip$ decays, as well as $\phi\to\Kp\Km$ decays in which a $\Kp$ meson is misidentified as a proton. These vetoes reduce the background in the $\Lb$ and $\Xibz$ samples by approximately 12\% each, while retaining about 98.5\% of the signal.

Signal $H_b$ candidates are divided into two subsamples based on the association of specific hardware triggers with reconstructed particles. The first category,
``triggered on signal'' (TOS), contains those candidates in which one or more of the final-state hadrons from the signal decay are associated with a positive L0 trigger decision from the calorimeter. The second category, ``triggered independently of the signal'' (TIS), represents the case where a positive L0 trigger decision is due to other particles in the event~\cite{LHCb-DP-2012-004}. For a given event, both the TOS and TIS requirements may be satisfied, in which case the event is classified as TOS. The TOS candidates tend to have larger $\pt$ than the TIS candidates due to the high energy transverse to the beamline that is required by the hadron trigger. Signal candidates are required to be in one of these two categories, which comprises 96\% of the $H_b$ signal decays selected by any L0 trigger in the data. At the level of the software trigger, the signal candidates are required to satisfy the requirements of the topological trigger, detailed in Refs.~\cite{BBDT,LHCb-PROC-2015-018}.

\subsection{Simulation corrections}
\label{sec:simcor}
\begin{figure}[tb]
\centering
\includegraphics[width=0.96\textwidth]{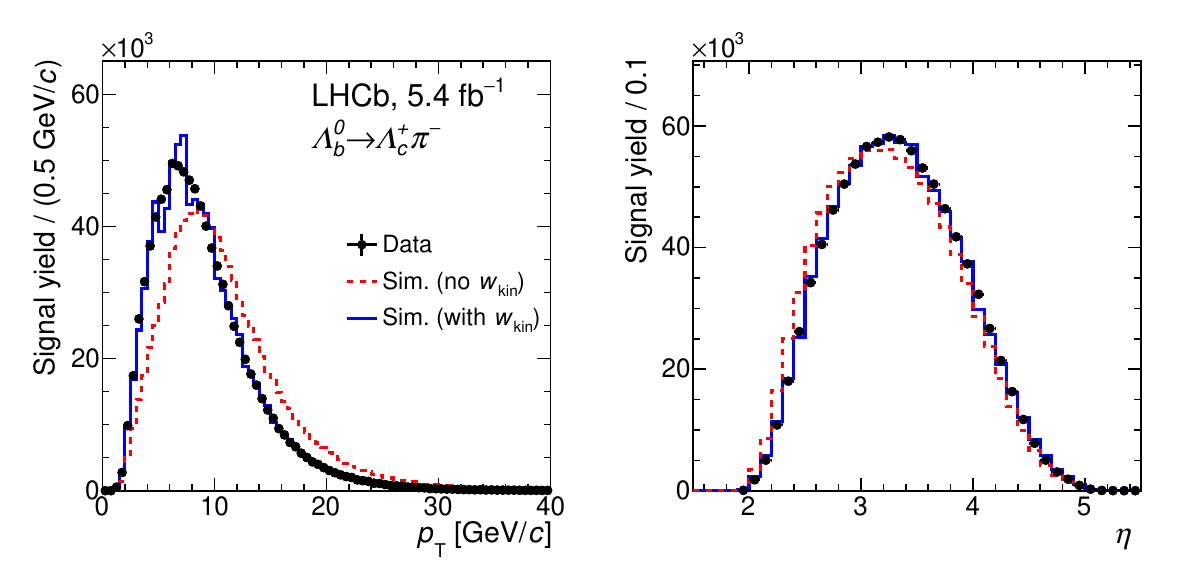}
\includegraphics[width=0.96\textwidth]{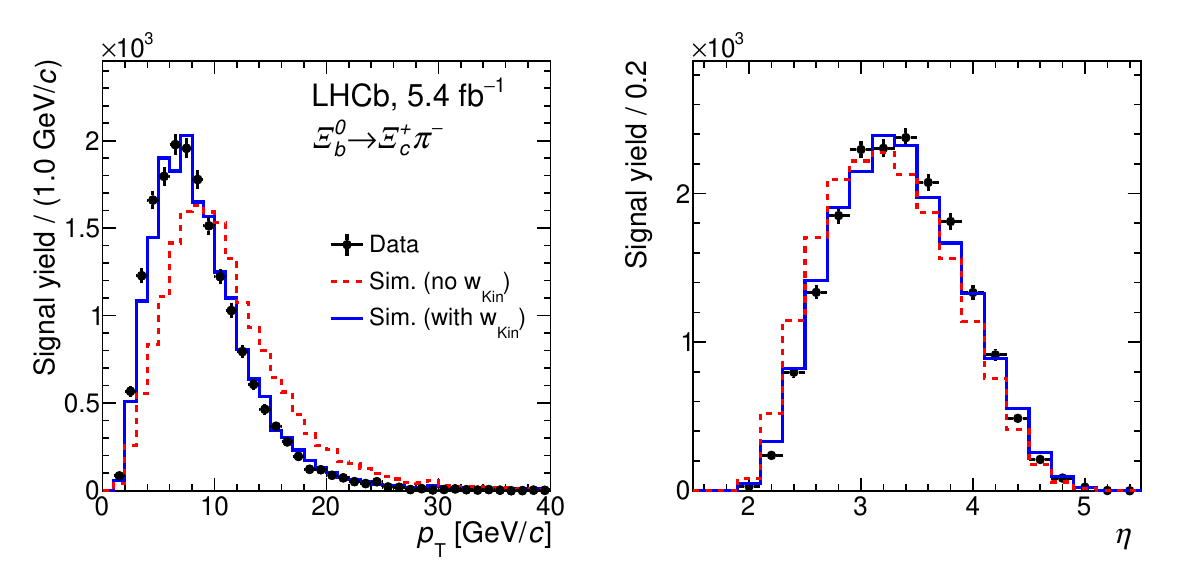}
\caption{\small{Comparison of the (left) {\ensuremath{p_{\mathrm{T}}}\xspace} and (right) $\eta$ spectra between background-subtracted data and simulation (Sim.), before (red dashed line), and after (blue line) the kinematic and TOS/TIS weights ($w_{\rm kin}$) are applied. The top row shows the distributions for the ${\it{\Lambda}}_b^0$ decay mode and the bottom shows those for the ${\it{\Xi}}_b^0$ mode.
}}
\label{fig:HbKinWeights}
\end{figure}

A set of weights and corrections are applied to the simulation to account for imperfect modeling of the signal decays and the detector response. The weights that are applied to the simulation correct for the $H_c$ decay kinematics, the $\Lb$ baryon lifetime, the $(\pt,\eta)$ spectra of the $H_b$ baryons and the fractions of TOS and TIS events. Weights to account for differences in the $H_c$ decay kinematics are obtained from the ratio of the two-dimensional mass distributions of $M(p\Km)$ versus $M(\Km\pip)$ in background-subtracted data relative to those obtained from simulation samples. The $\Lb$ lifetime weight accounts for a small difference between the value of $\tau_{\Lb}$ used in the simulation (1.451~ps) and the known value of 1.468~ps~\cite{PDG2024}. The $(\pt,\eta)$ weights ($w_{\rm kin}$) for $H_b$ are obtained by taking the ratio of the background-subtracted signal distributions in data to the corresponding distributions in simulation, separately for each L0 trigger category. Lastly, TIS/TOS weights are then applied to the simulation so that the fractions of TOS and TIS events match those in the data.  The $\pt$ and $\eta$ spectra of background-subtracted data~\cite{Pivk:2004ty} and simulation with and without the kinematic weights are compared in Fig.~\ref{fig:HbKinWeights}, showing improved agreement between the data and simulation after the correction is applied. 

\begin{figure}[tb]
\centering\includegraphics[width=0.96\textwidth]{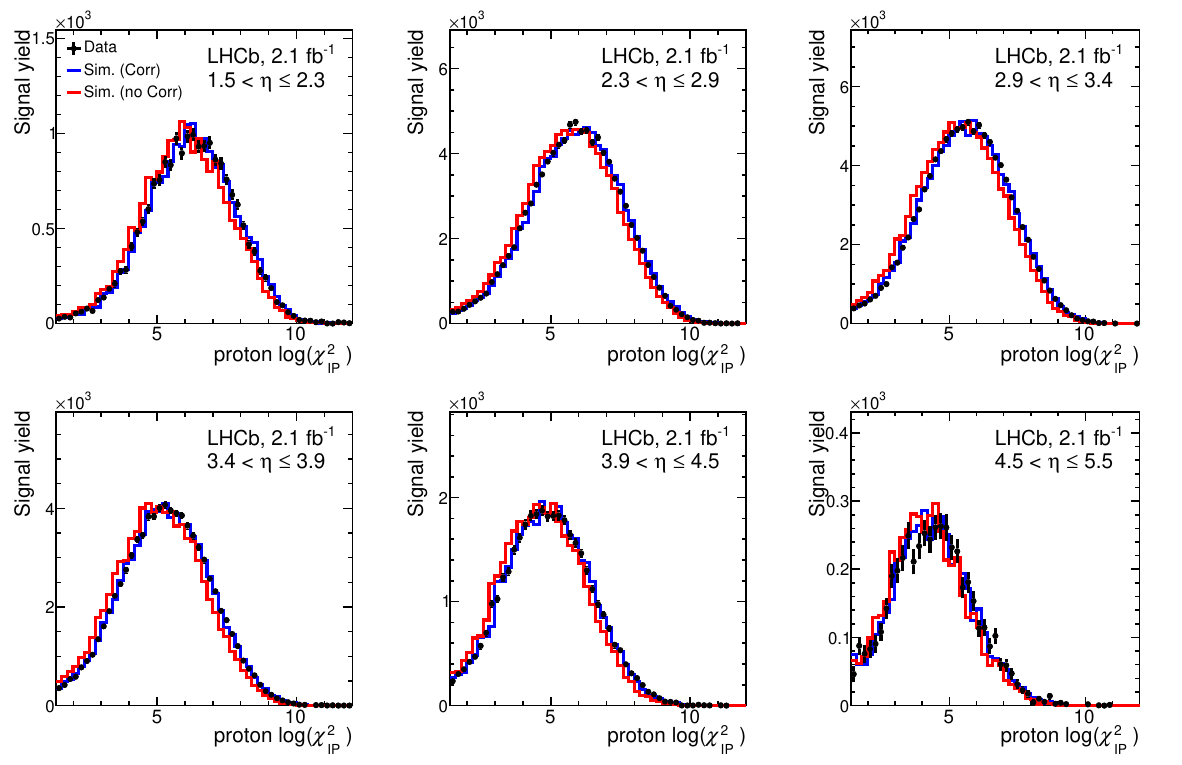}
\caption{\small{Distributions of $\log({\ensuremath{\chi^2_{\text{IP}}\xspace}})$ for background-subtracted 2018 data, uncorrected simulation (red line), and corrected simulation (blue line) for the proton from ${\it{\Lambda}}_b^0\to{\it{\Lambda}}_c^+\pi^-$ decays. The plots show the distributions in six $\eta$ regions of the proton.
}}
\label{fig:p_2018_vs_eta}
\end{figure}

In addition to the weights, a correction is applied to the 2017 and 2018 simulations to account for differences in the IP resolution and
in the $\chisq$ distributions from the vertex fit between data and simulation, as described in Ref.~\cite{LHCb-PAPER-2024-010}. Small adjustments are made to the $\chisqip$ corrections based on finer binning in $\eta$ and momentum to further improve the agreement, as shown in Fig.~\ref{fig:p_2018_vs_eta} for the proton from $\Lb\to\Lc\pim$ decays in 2018 data and simulation. A similar level of agreement is obtained for the kaon and pions. The same corrections are applied to the $\Xibz\to\Xicp\pim$ decay mode.

The PID response for the final-state hadrons is calibrated using large samples of \mbox{$\Dstarp\to\Dz(\Km\pip)\pip$} and \mbox{${\Lb\to\Lc\pim}$} decays~\cite{LHCb-DP-2018-001}, where no PID requirements are imposed. Using the calibration data, the PID response for each hadron type is parameterized as a function of its transverse and total momentum, as well as the track multiplicity in the event~\cite{Poluektov_2015}. The PID response of each hadron in simulated signal decays is updated to use the response obtained from these calibration data, sampled from the relevant PID probability distribution. All PID requirements applied to the simulated signal decays use these calibrated values.

\subsection{Multivariate selection}
To suppress background caused by random combinations of reconstructed particles, particularly in the $\Xibz$ decay mode, a gradient-boosted decision tree (BDT) classifier~\cite{AdaBoost,Hocker:2007ht} is employed. The training variables of the $\Lb$ ($\Xibz$) BDT classifier include the $H_b$ decay vertex-fit $\chisq$ and $\cos(\theta_{\rm PV})$, where $\theta_{\rm PV}$ is the angle between the $H_b$ momentum vector and the vector that connects the PV and $H_b$ decay vertex; the $H_c$ decay vertex-fit $\chisq$ and the $H_c$ baryon decay time; and for each final-state particle, the $p$, $\pt$, $\chisqip$ values and a PID variable providing a measure of the likelihood that the PID information is consistent with the particle hypothesis~\cite{LHCb-DP-2018-001}.

The signal distributions are taken from simulated signal decays with all weights applied.  The background sample for the $\Xibz$ ($\Lb$) mode is taken from the high-mass sideband region, \mbox{5900--6100\mevcc} \mbox{(5750--5950\mevcc)}, after all selection criteria are applied. For the $\Xibz$ decay, the $\Xicp$ mass requirement is relaxed to include candidates within 30\mevcc of the known $\Xicp$ mass~\cite{PDG2024} to increase the size of the background sample. A loose requirement is applied to the BDT algorithm output, which provides a signal efficiency of about 99\%, while suppressing the combinatorial background by a factor of four (three) for the $\Lb$ ($\Xibz$) decay mode. After the BDT requirement, less than 1\% of events have more than one $H_b$ candidate, and all candidates are retained~\cite{Koppenburg:2017zsh}.

\subsection{Fits to the mass distributions}
\begin{figure}[tb]
\centering
\includegraphics[width=0.48\textwidth]{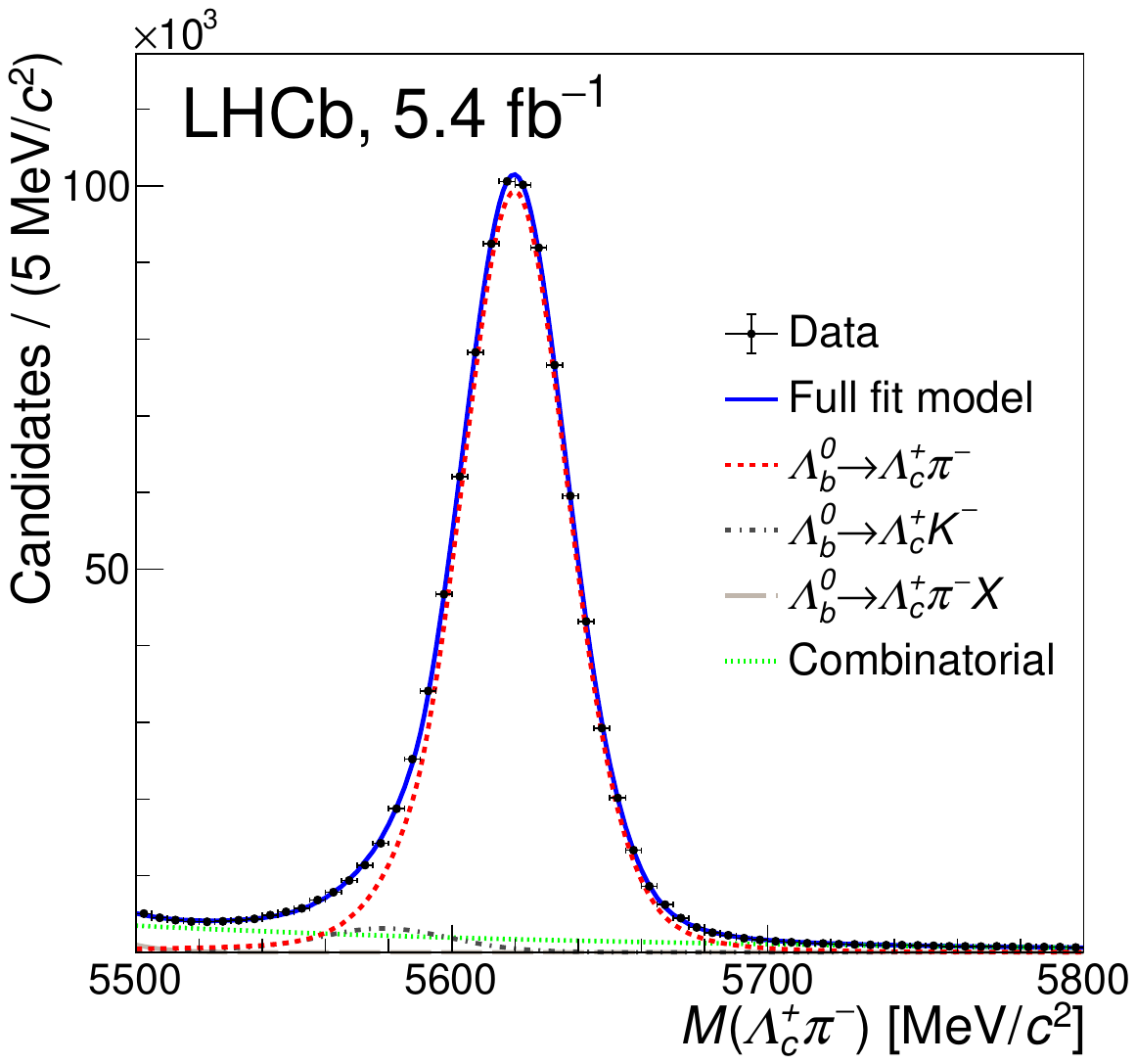}
\includegraphics[width=0.48\textwidth]{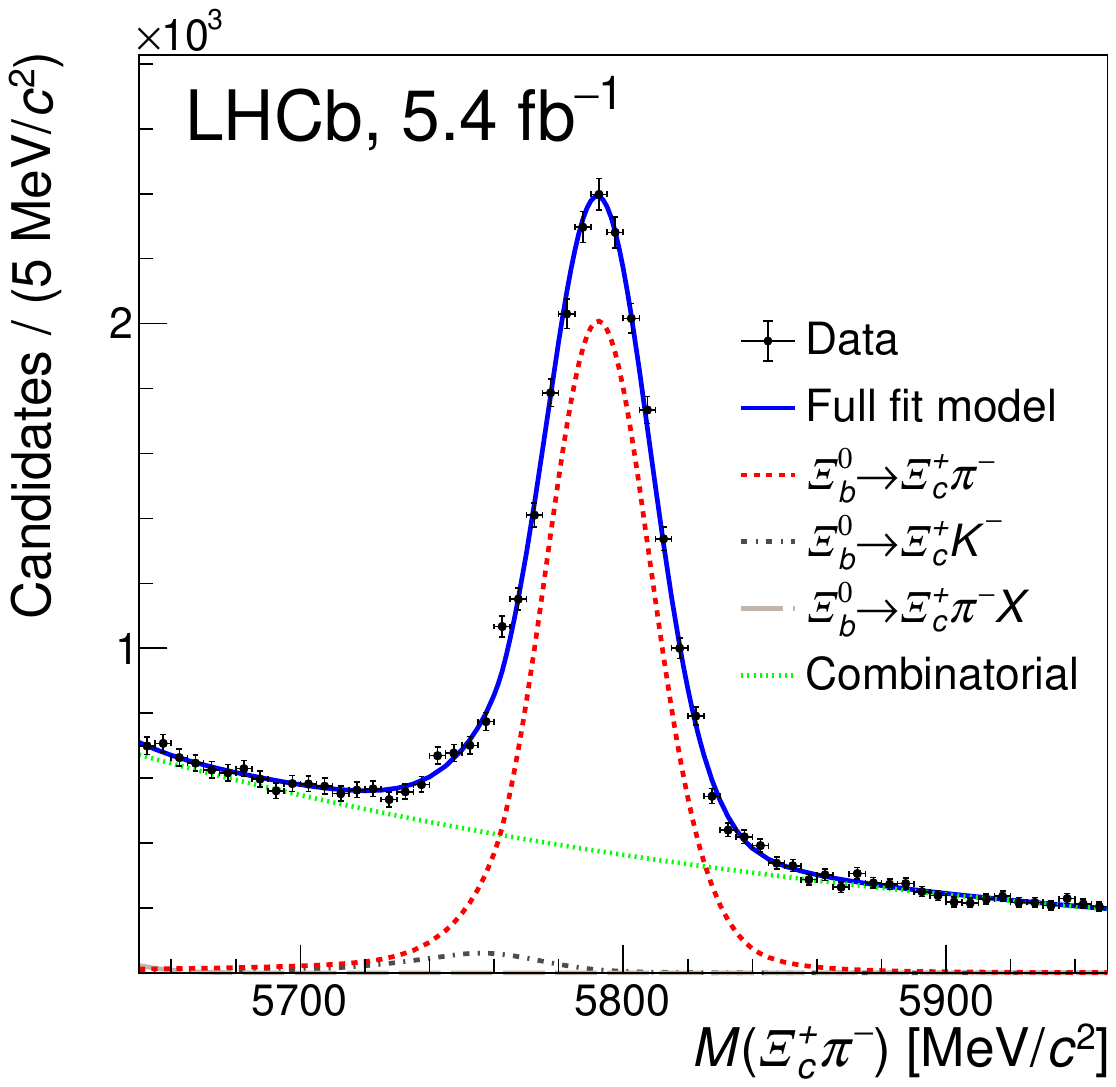}
\caption{\small{Mass spectra for (left) ${\it{\Lambda}}_b^0\to{\it{\Lambda}}_c^+\pi^-$ and (right) ${\it{\Xi}}_b^0\to{\it{\Xi}}_c^+\pi^-$ candidates in the data. The fit results and projections from the different components, as described in the text, are also shown.}} 
\label{fig:MassFitsFullSel}
\end{figure}

The mass spectra from the full sample for selected candidates are shown in Fig.~\ref{fig:MassFitsFullSel}, along with the results of binned extended maximum-likelihood fits~\cite{BARLOW1990496}. Each mass spectrum is described by the sum of a signal function and three background shapes. The signal-mass shapes are parameterized by the sum of two Crystal Ball functions~\cite{Skwarnicki:1986xj} with a common peak position. The signal shape parameters are fixed to the values obtained from simulation, except for the peak position and an overall scale factor, which accounts for a small difference in the mass resolution between data and simulation. In fits to the data, the resulting scale factor is about 1.10, which is close to the value obtained for the measurement of the $\Xibm$ baryon lifetime~\cite{LHCb-PAPER-2024-010}.

The mass shapes of misidentified $H_b\to H_c\Km$ decays are also described using the sum of two Crystal Ball functions~\cite{Skwarnicki:1986xj}, with shape parameters fixed to the values obtained from simulated $\Lb\to\Lc\Km$ decays. Taking into account the ratio of branching fractions 
\mbox{$\BR(\Lb\to\Lc\Km)/\BR(\Lb\to\Lc\pim)$~\cite{PDG2024}} and the relative selection efficiency, it is estimated that the yield fraction \mbox{$N(H_b\to H_c\Km)/N(H_b\to H_c\pim)$} is \mbox{$(3.1\pm0.6)\%$} for the $\Lb$ mode, and \mbox{$(3.1\pm0.9)\%$} for the $\Xibz$ decay. An additional 20\% relative uncertainty is included for the $\Xibz$ mode to account for an assumption that the relative branching fraction between the Cabibbo-suppressed and Cabibbo-favored mode is equal to that of the $\Lb$ baryon~\cite{PDG2024}. These fractions are constrained in the fit with Gaussian priors. Toward the lower edges of the mass distributions there are small contributions from partially reconstructed \bquark-hadron decays with a missing pion or photon(s). The partially reconstructed background is modeled with an Argus shape~\cite{ARGUS:1990jet} convolved with a Gaussian resolution function. The shape parameters are obtained from a fit to the data in a wider mass region, and then
fixed to those values in the default mass fits.
The combinatorial background is described by an exponential function with the shape parameter free to vary in the fit. The $\Xibz$ and $\Lb$ signal yields for each L0 category and both L0 categories combined are given in Table~\ref{tab:HbYieldsWithBDT}.

\begin{table*}[tb]
\begin{center}
\caption{\small{Signal yields by hardware trigger category for {{\ensuremath{\Lz^0_\bquark}}\xspace} and {{\ensuremath{\Xires^0_\bquark}}\xspace} decays for the total 2016--2018 dataset after all selection requirements. Each yield is obtained from an independent fit to the respective mass spectrum.}}
\begin{tabular}{lcccc}
\hline\hline
\\[-2.0ex]
$\bquark$ baryon         &   TOS  &   TIS   & TOS + TIS \\
\hline          
\\[-2.2ex]
$\Xibz$ ($\times10^3$)     &      $~\,10.1\pm0.1$     &   $~~\,8.1\pm0.1$  &   $~\,18.1\pm0.2$ \\
\\[-2.2ex]
\hline
\\[-2.2ex]
$\Lb$ ($\times10^3$) &    $520.7\pm0.9$   &  $408.2\pm0.8$ &  $929.1\pm1.2$ \\
\\[-2.2ex]
\hline\hline
\end{tabular}
\label{tab:HbYieldsWithBDT}
\end{center}
\end{table*}

\section{\boldmath Determination of $r_{\tau}$}
\begin{figure}[tb]
\centering
\includegraphics[width=0.48\textwidth]{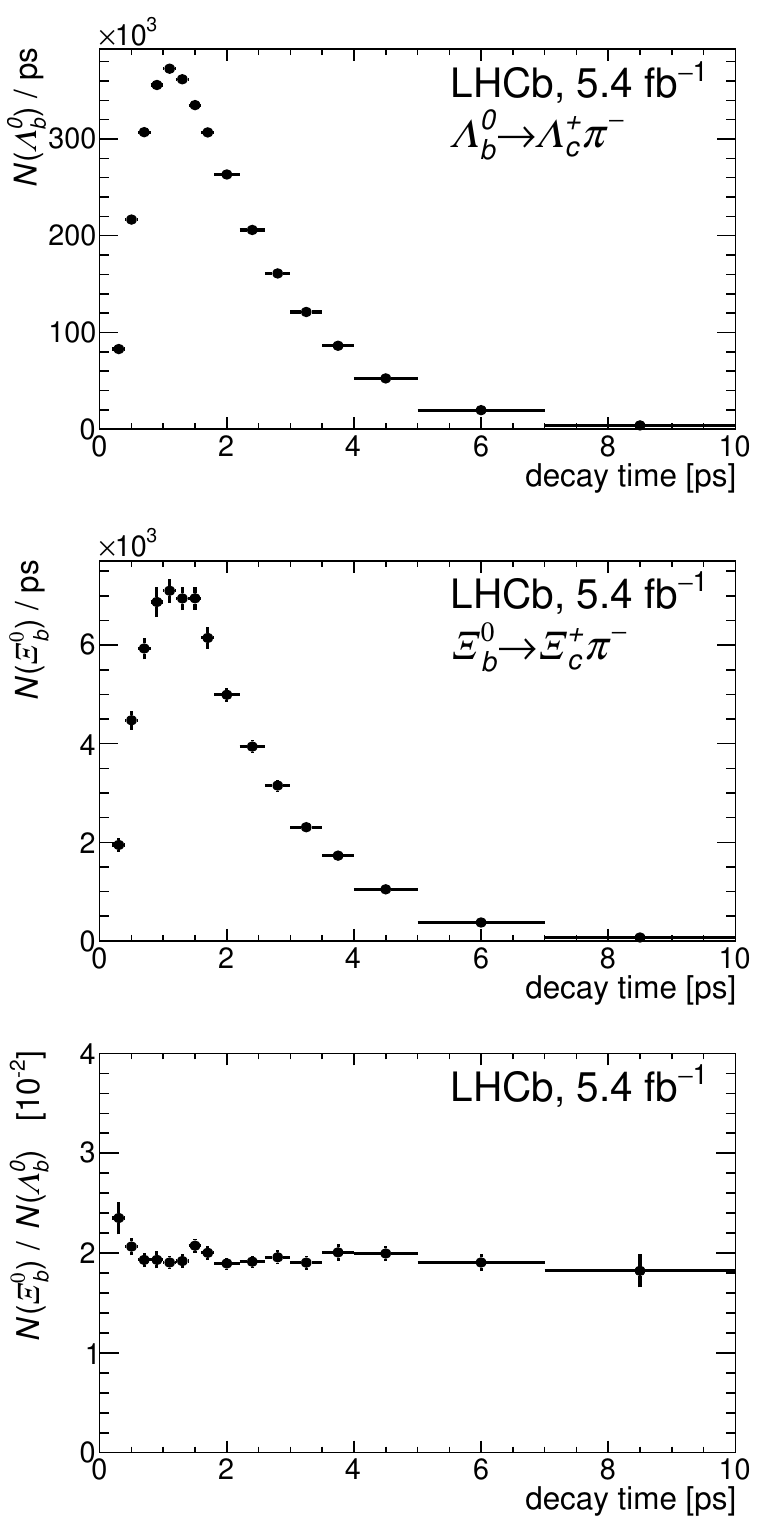}
\includegraphics[width=0.48\textwidth]{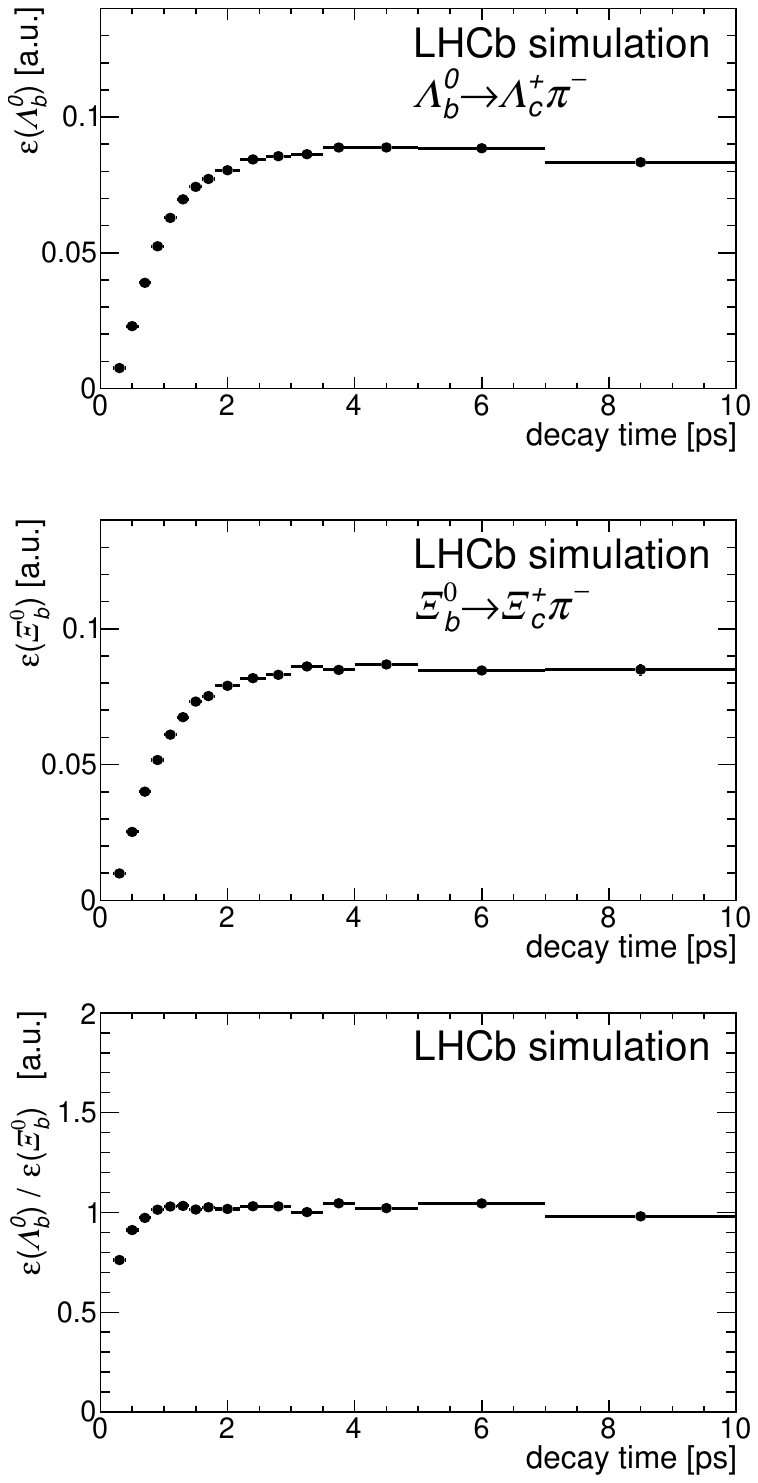}
\caption{\small{(Left) Decay time dependence of the signal yields divided by the bin width for the (top) ${\it{\Lambda}}_b^0$ mode, (middle) ${\it{\Xi}}_b^0$, and (bottom) the ratio $N({\it{\Xi}}_b^0)/N({\it{\Lambda}}_b^0)$. (Right) Decay time dependence of the selection efficiencies for the (top) ${\it{\Lambda}}_b^0$ mode, (middle) ${\it{\Xi}}_b^0$ mode, and (bottom) the ratio $\varepsilon({\it{\Lambda}}_b^0)/\varepsilon({\it{\Xi}}_b^0)$, obtained from the weighted simulation. The uncertainties shown are due to the finite number of events in the data and simulation samples.}}
\label{fig:RawYieldsSum}
\end{figure}

The determination of the lifetime ratio $r_{\tau}$ requires the ratio of signal yields and the ratio of selection efficiencies, as shown in Eq.~\ref{eq:RT}.
The yields \mbox{$N[\Xibz\to\Xicp\pim](t)$} and \mbox{$N[\Lb\to\Lc\pim](t)$}
are determined by fitting the combined TOS+TIS $H_b$ mass spectra in decay-time bins, ranging from 0.4 to $10.0\ps$. Specifically, the bin widths are $0.2\ps$ from 
$[0.4,\,1.8]\ps$, $0.4\ps$ from $[1.8,\,3.0]\ps$, $0.5\ps$ from $[3.0,\,4.0]\ps$,
and one bin each from [4.0,\,5.0], [5.0,\,7.0] and $[7.0,\,10.0]\ps$. The mass fits are performed as described previously, where the signal shape parameters in each decay-time bin are obtained from the corresponding fit to the simulated signal decays in the same decay-time bin, with all weights applied. For the $\Xibz$ mode, the mass-resolution scale factor is constrained by a Gaussian prior to the value obtained from the fit to the full data sample. The independence of the decay-time resolution on decay time is validated with simulation. All signal and background yields are free parameters in the fit for each decay-time bin. The yields of $\Lb$ and $\Xibz$ baryon decays as a function of the decay time are shown in Fig.~\ref{fig:RawYieldsSum}\,(left), along with the ratio of yields, \mbox{$N(\Xibz)/N(\Lb)$}. 

The selection efficiencies for decays within the LHCb acceptance are obtained from simulation, corrected for data-simulation differences as detailed in Section~\ref{sec:simcor}. The resulting efficiencies versus decay time are shown in Fig.~\ref{fig:RawYieldsSum}\,(right). The $\Lb$ and $\Xibz$ signal efficiencies show a steep rise from zero decay time, primarily due to the $\chisqip$ requirements in the trigger and selection. For the lifetime determination, the $H_b$ decay time is required to be larger than 0.4~ps. As shown in the bottom plots of Fig.~\ref{fig:RawYieldsSum}, the increase in the yield ratio $N(\Xibz)/N(\Lb)$ at low decay time trends in the opposite direction to the ratio of selection efficiencies $\eff(\Lb)/\eff(\Xibz)$, as one would expect based on the longer $\Xicp$ baryon lifetime compared to that of the $\Lc$ baryon.

\begin{figure}[tb]
\centering
\includegraphics[width=0.80\textwidth]{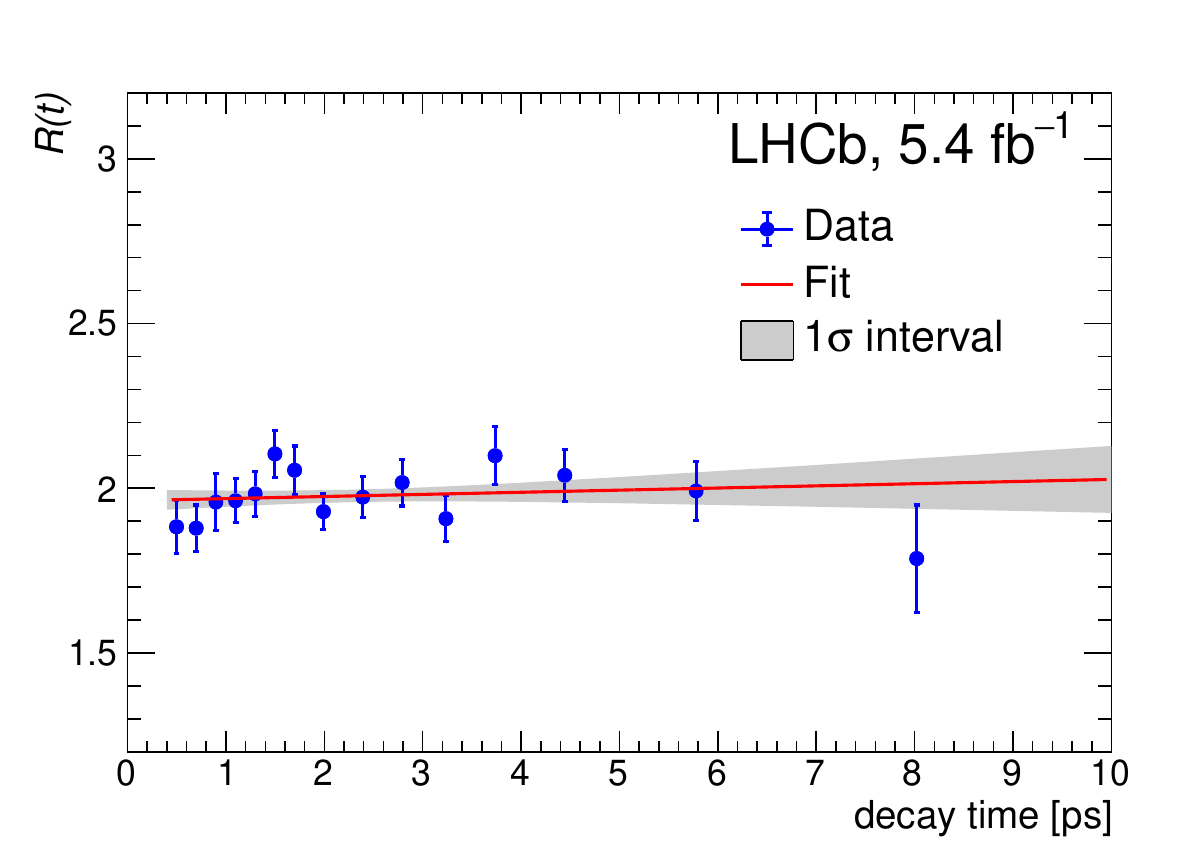}
\includegraphics[width=0.80\textwidth]{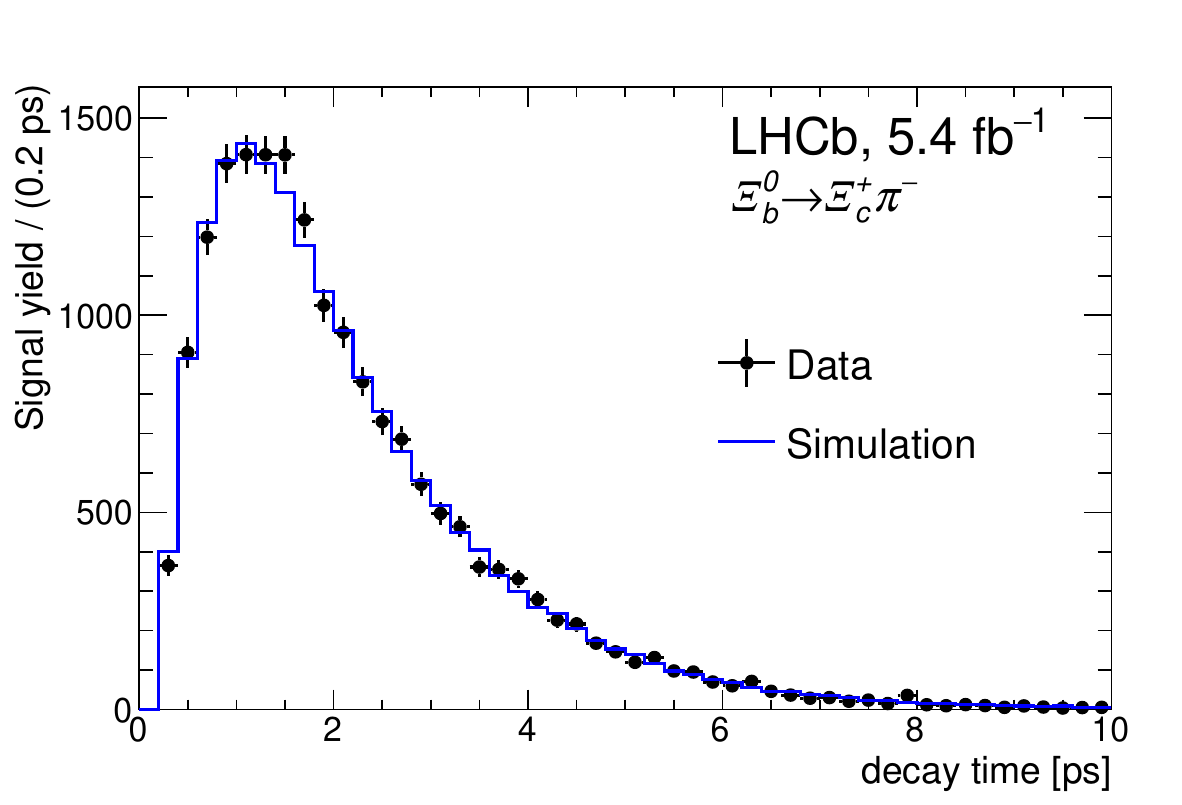}
\caption{\small{(Top) Efficiency-corrected yield ratio $R(t)$ as a function of the decay time. The red line shows the fit result, and the gray band shows the confidence interval at one standard deviation ($\sigma$). The points are placed along the time axis at the average within the bin~\cite{Lafferty}, assuming a decay-time spectrum that is exponential with an effective lifetime of 1.47~ps. (Bottom) Decay-time spectrum of ${\it{\Xi}}_b^0$ signal decays, along with the simulation using the best-fit lifetime of 1.473~ps.
}}
\label{fig:CorrYieldRatio}
\end{figure}

The product of the yield and efficiency ratios shown in Fig.~\ref{fig:RawYieldsSum}\,(bottom) gives $R(t)$, and the result is shown in Fig.~\ref{fig:CorrYieldRatio}\,(top). The results of a fit to an exponential function are also shown, yielding 
\mbox{$\lambda=(2.5\pm6.4)\times10^{-3}\ps^{-1}$}, with a $\chisq$ per degree of freedom of 1.04. The uncertainties shown are due to the finite signal yields in the data. From this value, the relative lifetime and $\Xibz$ lifetime are computed to be
\begin{align}
    r_{\tau}&=1.004\pm0.009, \nonumber\\
    \tau_{\Xibz}&=1.473\pm0.014~{\rm ps},\nonumber
\end{align}
\noindent where the uncertainties are statistical only.

Figure~\ref{fig:CorrYieldRatio}\,(bottom) compares the $\Xibz$ baryon decay-time spectrum in background-subtracted data to that of the simulation weighted to represent the measured $\Xibz$ lifetime. The simulation using the fitted lifetime provides a good description of the spectrum observed in the data.

\section{Cross-checks and systematic uncertainties}
A number of cross-checks are performed, where the data are separated into statistically independent subsamples, and the $r_{\tau}$ value is measured in each of these subsamples. The subsamples are partitioned according to (a) data-taking year, (b) TOS and TIS, (c) polarity of the LHCb magnet, (d) number of tracks per event ($<150$ or $\geq150$), and (e) number of primary vertices ($n$PV$\,=\,$1 or 
$n$PV\,$\geq\,$2). In all cases, the $r_{\tau}$ values are statistically consistent with each other. Among the 11 subsamples, the largest deviation from the overall average is about $1.2\sigma$.

Several sources of systematic uncertainty are considered and shown in Table~\ref{tab:Syst}, along with their associated values. For each of the sources, unless otherwise indicated, the systematic uncertainty is assessed by making a change to the default method, and assigning the relative change in $r_{\tau}$ with respect to the baseline value as the systematic uncertainty. The total systematic uncertainty is obtained from the quadrature sum of the values from each individual source.

\begin{table*}[tb]
\begin{center}
\caption{\small{Sources and values of the relative systematic uncertainty on the lifetime ratio measurement.}}
\begin{tabular}{lc}
\hline\hline
Source                    &   Value (\%) \\
\hline
Simulated sample size     &     0.31     \\
Signal shape              &     0.28     \\  
Background shape          &     0.02     \\
$\Xibz\to\Xicprime\pim$ background    &     0.15     \\  
Truth matching            &     0.03     \\    
Bin width in mass         &     0.10     \\  
Bin width in time         &     0.06     \\   
BDT requirement           &     0.36     \\  
$\Lb$ lifetime            &     0.00 \\
\hline
Total                     &     0.58 \\
\hline\hline
\end{tabular}
\label{tab:Syst}
\end{center}
\end{table*}

The systematic uncertainty due to the finite simulated sample sizes is obtained by performing 1000 alternative fits for $r_{\tau}$, where in each fit the relative efficiency in each bin is fluctuated by its uncertainty according to a Gaussian distribution. The standard deviation of the 1000 alternative fits is assigned as the systematic uncertainty.

The sensitivity to the signal shape model is assessed by using an alternative signal model composed of the sum of a Bukin~\cite{Verkerke:2003ir} and a Gaussian function with a common peak value. As in the baseline fit, the parameters are determined from simulation and fixed in the mass fits, apart from a scaling of the resolution and the peak mass value. Similarly,
the uncertainty due to a specific choice for the combinatorial background model is estimated by redoing the analysis using a third-order (second-order) Chebyshev polynomial for the $\Lb$ ($\Xibz$) mass fits instead of an exponential function. 

In the baseline fit, background from the unobserved decay \mbox{${\Xibz\to\Xicprime\pim}$} with \mbox{$\Xicprime\to\Xicp\gamma$}, has been neglected. In one model calculation~\cite{ke2024studymixingxicxic}, the ratio of partial widths is small, \mbox{$\Gamma(\Xib\to\Xicprime\pi)/\Gamma(\Xib\to\Xic\pi)\simeq6.3\%$}.
The fractional contribution is taken to be
$(5.3\pm0.6)\%$, which includes a correction for the fraction of decays in the fit range. 
The shape of this partially reconstructed background is parameterized using {\textsc{RapidSim}}~\cite{Cowan:2016tnm} and included in the mass fits with the fractional contribution constrained by Gaussian prior to the value above. The newly obtained $\Xibz$ signal yields lead to a modified value of $r_{\tau}$, and the relative change from the baseline result is assigned as the systematic uncertainty.

For the efficiency determination, the reconstructed signal decays in the simulation are required to be truth-matched with the generated signal decays. In about 7\% of cases, this association fails due to having too few hits on the reconstructed tracks that match to the true hits of each final-state particle. The potential impact of the truth-matching requirement is assessed by removing the matching requirement entirely for the efficiency determination.

The fits are binned extended maximum-likelihood fits, and the baseline bin width is 5.0\mevcc for both the $\Lb$ and $\Xibz$ mass fits. In an alternative set of fits, the corresponding bin widths are reduced to 1.0\mevcc and 2.5\mevcc,
the fits are redone, and the relative difference is assigned as a systematic uncertainty. To study the effects of the binning in decay time, the number of decay-time bins is increased from 17 to 24. The mass fits in the narrower decay-time bins are repeated, leading to a new set of data points for $R(t)$. The relative change from the baseline result is assigned as a systematic uncertainty.

The BDT performances in data and simulation do not match perfectly for the signal and normalization modes. Since the BDT uses decay-time biasing variables, mainly $\chisqip$, differences in the BDT output could be indicative of an imperfect modeling of such input variables and their correlations with other input variables. To assess the potential impact, the ratio of the BDT response in background-subtracted data~\cite{Pivk:2004ty} to simulation is obtained, and used as an additional weight in the determination of the efficiencies \mbox{$\eff[H_b\to H_c\pim](t)$}. The resulting fractional change in $r_{\tau}$ is assigned as a systematic uncertainty.

Due to the small value of $\lambda$ and the precisely known $\Lb$ baryon lifetime, 
{\mbox{$\tau_{\Lb}=1.468\pm0.009$ ps}~\cite{PDG2024}}, the systematic uncertainty in $r_{\tau}$ is
negligible. The total systematic uncertainty on $r_{\tau}$ is 0.58\%, which is subdominant compared to the statistical precision of 0.94\%.

\section{Results and Summary}
From a $pp$ collision data sample corresponding to an integrated luminosity of 5.4\invfb, the LHCb collaboration measures the ratio of lifetimes of the $\Xibz$ baryon relative to that of the $\Lb$ baryon to be
\begin{align*}
    r_{\tau}^{\rm Run\,2} = 1.004\pm0.009\pm0.006,
\end{align*}
\noindent where the uncertainties are statistical and systematic, respectively. Multiplying this ratio by the known $\Lb$ baryon lifetime, the $\Xibz$ baryon lifetime is found to be
\begin{align*}
    \tau_{\Xibz}^{\rm Run\,2} = 1.473\pm0.014\pm0.009\pm0.009~{\rm ps},
\end{align*}
\noindent where the last uncertainty is due to that of the $\Lb$ baryon lifetime. 

The value of $r_{\tau}$ is consistent with and about two times more precise than the value $1.006\pm0.018\pm0.010$ obtained using Run 1 data~\cite{LHCb-PAPER-2014-021}. The two results are combined taking only the finite sample sizes of the simulated samples as uncorrelated. The rest of the systematic uncertainties are taken to be 100\% correlated. Using the BLUE package~\cite{Nisius:2020jmf},
the Run 1--2 average values are
\begin{align*}
r_{\tau} &= 1.004\pm0.008\pm0.005,\\
\tau_{\Xibz} &= 1.475\pm0.012\pm0.008\pm0.009~{\rm ps}.
\end{align*}
This corresponds to a precision on the lifetime ratio of 1\%, which represents an improvement on the Run 1 value by about a factor of two. The measured value is in agreement with the prediction from the heavy-quark expansion, of \mbox{$r_{\tau}^{\rm HQE}=1.002\pm0.023$~\cite{Gratrex}}. This prediction only accounts for the decay width of the $\bquark$ quark and higher-order corrections. Decays of the $\Xibm$ baryon involving the weak $s\to u\bar{u}d$ transition have been observed by the LHCb collaboration,
with branching fraction \mbox{$\BR(\Xibm\to\Lb\pim)\sim1\%$~\cite{LHCb-PAPER-2023-015}}. Similar decays are expected for the $\Xibz$ baryon~\cite{Li:2014ada,VOLOSHIN2000297}, which would reduce the central value of the theoretical prediction by about 0.5\%. After applying this correction the theoretical calculation would still be in agreement with the measured $r_{\tau}$ value.

\section*{Acknowledgements}
%
%
\noindent We express our gratitude to our colleagues in the CERN
accelerator departments for the excellent performance of the LHC. We
thank the technical and administrative staff at the LHCb
institutes.
We acknowledge support from CERN and from the national agencies:
ARC (Australia);
CAPES, CNPq, FAPERJ and FINEP (Brazil); 
MOST and NSFC (China); 
CNRS/IN2P3 (France); 
BMBF, DFG and MPG (Germany); 
INFN (Italy); 
NWO (Netherlands); 
MNiSW and NCN (Poland); 
MCID/IFA (Romania); 
MICIU and AEI (Spain);
SNSF and SER (Switzerland); 
NASU (Ukraine); 
STFC (United Kingdom); 
DOE NP and NSF (USA).
We acknowledge the computing resources that are provided by ARDC (Australia), 
CBPF (Brazil),
CERN, 
IHEP and LZU (China),
IN2P3 (France), 
KIT and DESY (Germany), 
INFN (Italy), 
SURF (Netherlands),
Polish WLCG (Poland),
IFIN-HH (Romania), 
PIC (Spain), CSCS (Switzerland), 
and GridPP (United Kingdom).
We are indebted to the communities behind the multiple open-source
software packages on which we depend.
Individual groups or members have received support from
Key Research Program of Frontier Sciences of CAS, CAS PIFI, CAS CCEPP, 
Fundamental Research Funds for the Central Universities,  and Sci.\ \& Tech.\ Program of Guangzhou (China);
Minciencias (Colombia);
EPLANET, Marie Sk\l{}odowska-Curie Actions, ERC and NextGenerationEU (European Union);
A*MIDEX, ANR, IPhU and Labex P2IO, and R\'{e}gion Auvergne-Rh\^{o}ne-Alpes (France);
Alexander-von-Humboldt Foundation (Germany);
ICSC (Italy); 
Severo Ochoa and Mar\'ia de Maeztu Units of Excellence, GVA, XuntaGal, GENCAT, InTalent-Inditex and Prog.~Atracci\'on Talento CM (Spain);
SRC (Sweden);
the Leverhulme Trust, the Royal Society and UKRI (United Kingdom).



\newpage

\addcontentsline{toc}{section}{References}
\bibliographystyle{LHCb}
\bibliography{main,standard,LHCb-PAPER,LHCb-CONF,LHCb-DP,LHCb-TDR}

\newpage
\centerline
{\large\bf LHCb collaboration}
\begin
{flushleft}
\small
R.~Aaij$^{38}$\lhcborcid{0000-0003-0533-1952},
A.S.W.~Abdelmotteleb$^{57}$\lhcborcid{0000-0001-7905-0542},
C.~Abellan~Beteta$^{51}$\lhcborcid{0009-0009-0869-6798},
F.~Abudin{\'e}n$^{57}$\lhcborcid{0000-0002-6737-3528},
T.~Ackernley$^{61}$\lhcborcid{0000-0002-5951-3498},
A. A. ~Adefisoye$^{69}$\lhcborcid{0000-0003-2448-1550},
B.~Adeva$^{47}$\lhcborcid{0000-0001-9756-3712},
M.~Adinolfi$^{55}$\lhcborcid{0000-0002-1326-1264},
P.~Adlarson$^{85}$\lhcborcid{0000-0001-6280-3851},
C.~Agapopoulou$^{14}$\lhcborcid{0000-0002-2368-0147}
% ...

{\footnotesize \it

% ...
$^{14}$Universit{\'e} Paris-Saclay, CNRS/IN2P3, IJCLab, Orsay, France\\
% ...
$^{38}$Nikhef National Institute for Subatomic Physics, Amsterdam, Netherlands\\
% ...
$^{47}$Instituto Galego de F{\'\i}sica de Altas Enerx{\'\i}as (IGFAE), Universidade de Santiago de Compostela, Santiago de Compostela, Spain\\
% ...
$^{51}$Physik-Institut, Universit{\"a}t Z{\"u}rich, Z{\"u}rich, Switzerland\\
% ...
$^{55}$H.H. Wills Physics Laboratory, University of Bristol, Bristol, United Kingdom\\
% ...
$^{57}$Department of Physics, University of Warwick, Coventry, United Kingdom\\
% ...
$^{60}$School of Physics and Astronomy, University of Glasgow, Glasgow, United Kingdom\\
$^{61}$Oliver Lodge Laboratory, University of Liverpool, Liverpool, United Kingdom\\
% ...
$^{69}$Syracuse University, Syracuse, NY, United States\\
% ...
$^{85}$Department of Physics and Astronomy, Uppsala University, Uppsala, Sweden, associated to $^{60}$\\
% ...
}
\end{flushleft}

\end{document}